\documentclass[8.5pt,twoside,twocolumn]{article}
\usepackage[super,sort&compress,comma]{natbib}
\usepackage[version=3]{mhchem}
\usepackage{balance}
\usepackage{times,mathptm}
\usepackage{sectsty}
\usepackage{graphicx}
\usepackage{lastpage}
\usepackage[format=plain,justification=raggedright,singlelinecheck=false,font=small,labelfont=bf,labelsep=space]{caption}
\usepackage{fancyhdr}
\begin{document}

\thispagestyle{plain}
\fancypagestyle{plain}{
\renewcommand{\headrulewidth}{1pt}}
\renewcommand{\thefootnote}{\fnsymbol{footnote}}
\renewcommand\footnoterule{\vspace*{1pt}%
\hrule width 3.4in height 0.4pt \vspace*{5pt}}
\setcounter{secnumdepth}{5}

\makeatletter
\renewcommand\@biblabel[1]{#1}
\renewcommand\@makefntext[1]%
{\noindent\makebox[0pt][r]{\@thefnmark\,}#1}
\makeatother
\renewcommand{\figurename}{\small{Fig.}~}
\sectionfont{\large}
\subsectionfont{\normalsize}

\fancyfoot{}
\fancyfoot[LE]{\footnotesize{\sffamily{\thepage~\textbar\hspace{3.45cm} 1--\pageref{LastPage}}}}
\fancyhead{}
\renewcommand{\headrulewidth}{1pt}
\renewcommand{\footrulewidth}{1pt}
\setlength{\arrayrulewidth}{1pt}
\setlength{\columnsep}{6.5mm}
\setlength\bibsep{1pt}

\twocolumn[
  \begin{@twocolumnfalse}
\noindent\LARGE{\textbf{Dichotomic Aging Behaviour in a Colloidal
Glass}} \vspace{0.6cm}

\noindent\large{\textbf{Roberta
Angelini,$^{\ast}$\textit{$^{a,b}$} Laura Zulian,\textit{$^{b}$}
Andrei Fluerasu, \textit{$^{c}$} Anders Madsen, \textit{$^{d}$}
Giancarlo Ruocco, \textit{$^{a,b,e}$} and Barbara Ruzicka
\textit{$^{a,b}$}}}\vspace{0.5cm}


 \end{@twocolumnfalse} \vspace{0.6cm}

  ]

\noindent\textbf{An unexpected dichotomic long time aging
behaviour is observed in a glassy colloidal clay suspension
investigated by X-Ray Photon Correlation Spectroscopy and Dynamic
Light Scattering. In the long time aging regime the intensity
autocorrelations are non-exponential, following the
Kohlrausch-Williams-Watts functional form with exponent $\beta_Q$.
We show that for spontaneously aged samples a stretched behaviour
($\beta_Q <1$) is always found. Surprisingly a compressed exponent
($\beta_Q>1$) appears only when the system is rejuvenated by
application of a shear field. In both cases the relaxation times
scale as $Q^{-1}$. These observations shed light on the origin of
compressed exponential behaviour and helps in classifying previous
results in the literature on anomalous dynamics.}

\section*{}
\vspace{-1cm}

\footnotetext{\textit{$^{a}$~CNR-IPCF, I-00185 Rome, Italy Fax:
+39064463158; Tel: +390649913433; E-mail:
roberta.angelini@roma1.infn.it}}
\footnotetext{\textit{$^{b}$Dipartimento di Fisica, Sapienza
Universit$\grave{a}$ di Roma, I-00185, Italy}}
\footnotetext{\textit{$^{c}$Brookhaven National Laboratory,
NSLS-II Upton NY 11973, USA}} \footnotetext{\textit{$^{d}$
European X-Ray Free-Electron Laser, Albert-Einstein-Ring 19,
D-22761 Hamburg, Germany }}
\footnotetext{\textit{$^{e}$ Istituto
Italiano di Tecnologie, Center for Life Nano Science, Sapienza,
Rome, Italy}}

The continuous interest in studying dynamics of colloidal systems
has over the last decade led to a broadening of our knowledge on
dynamical arrest and the glass transition. Many experiments
performed on colloidal glasses~\cite{PuseyJPCM2008} and
gels~\cite{Trappe2004, Poon1998} as well as theoretical and
numerical simulation~\cite{ZaccarelliJPCM2007, SciortinoAP2005}
aim at understanding the nature of the aging phenomena that are
typical for systems with time evolving dynamics. Studying the time
evolution of the dynamic structure factor is the most direct way
to access microscopic information on aging. It allows extracting
the characteristic times of the system (relaxation times) as well
as their distribution ($\beta_Q$ exponent, as described below).

The commonly accepted physical picture in a disordered,
glass--like material, is that the motion of a particle is
constrained by the cage of its neighbors resulting  in the
emergence of different relaxation times - e.g. relation to the
motion inside the cage, rearrangements of cages, etc. In order to
fit the decay of the intensity correlation functions over a wide
time window, the Kohlrausch-Williams-Watts expression is generally
used $f(Q,t)\sim exp[-(t/\tau_Q)^{\beta_Q}]$ where $\tau_Q$ is an
``effective'' relaxation time and $\beta_Q$ measures the
distribution of relaxation times (associated with simple
exponential decays). Most commonly, the different relaxation times
present in glassy materials lead to a stretching of the
correlation functions and an exponent $\beta_Q<1$ (which here is
referred to as ``stretched behaviour'').

On the contrary, the pioneering work of Cipelletti {\em et
al.}~\cite{CipellettiPRL2000} on a gel of colloidal polystyrene
has shown the existence of anomalous dynamics. A peculiar aging
behavior has been identified with exponential growth of the
characteristic relaxation times at early ages $t_w$ and an almost
linear trend at larger $t_w$ ($\tau_Q \sim$ $t_w^{\alpha}$ with
$\alpha = 0.9 \pm 0.1$). Moreover, $\tau_Q \sim Q^{-1}$ with
$\beta_Q>1$ (compressed behaviour) is found. This unusual dynamics
has been attributed to the relaxation of internal
stresses~\cite{CipellettiFD2003}.

More recently, this novel anomalous dynamics, fully investigated
through X-ray Photon Correlation Spectroscopy (XPCS) in a wide
wave vector region, has been recognized as a salient feature of
disordered arrested materials ranging from glassy polymers and
supercooled molecular liquids to soft matter
systems~\cite{MadsenNJP2010,LehenyCOCIS2012}. Among these, a
colloidal clay (Laponite) has recently emerged as complex fluid
system characterized by a peculiar aging
dynamics~\cite{BellourPRE2003,
BandyopadhyayPRL2004,TanakaPRE2005,SchosselerPRE2006}.

The Dynamic Light Scattering (DLS) work of Bellour {\em et
al.}~\cite{BellourPRE2003} on Laponite, reports an exponential
increase of $\tau_Q$ at small $t_w$ and a power law behaviour with
$\alpha = 1.0 \pm 0.1$ accompanied by $\beta_Q>1$ at larger $t_w$,
in agreement with Ref.~\cite{CipellettiPRL2000}. Multi-speckle
XPCS measurements performed by Bandyopadhyay {\em et
al.}~\cite{BandyopadhyayPRL2004} on Laponite suspensions showed an
hyper-diffusive dynamics ($\beta_Q>1$, $\tau \sim Q^{-1}$) and a
power law behaviour of the relaxation time $\tau_Q\sim
t_w^{\alpha}$ with $\alpha = 1.8 \pm 0.2$ in the large $t_w$
range. Subsequently, Tanaka {\em et al.}~\cite{TanakaPRE2005}
classified these two distinct dynamical aging regimes as
cage-forming at small $t_w$ and full aging at larger $t_w$.
Finally, Schosseler {\em et al.}~\cite{SchosselerPRE2006}
confirmed the existence of these two dynamical regimes reporting
that the crossover between them depends on wave vector and that
the full aging is characterized by $\alpha = 1.04 \pm 0.06$ and
$\beta_Q>1$.

\begin{figure}
\centering
\includegraphics[width=7.5cm,angle=0,clip]{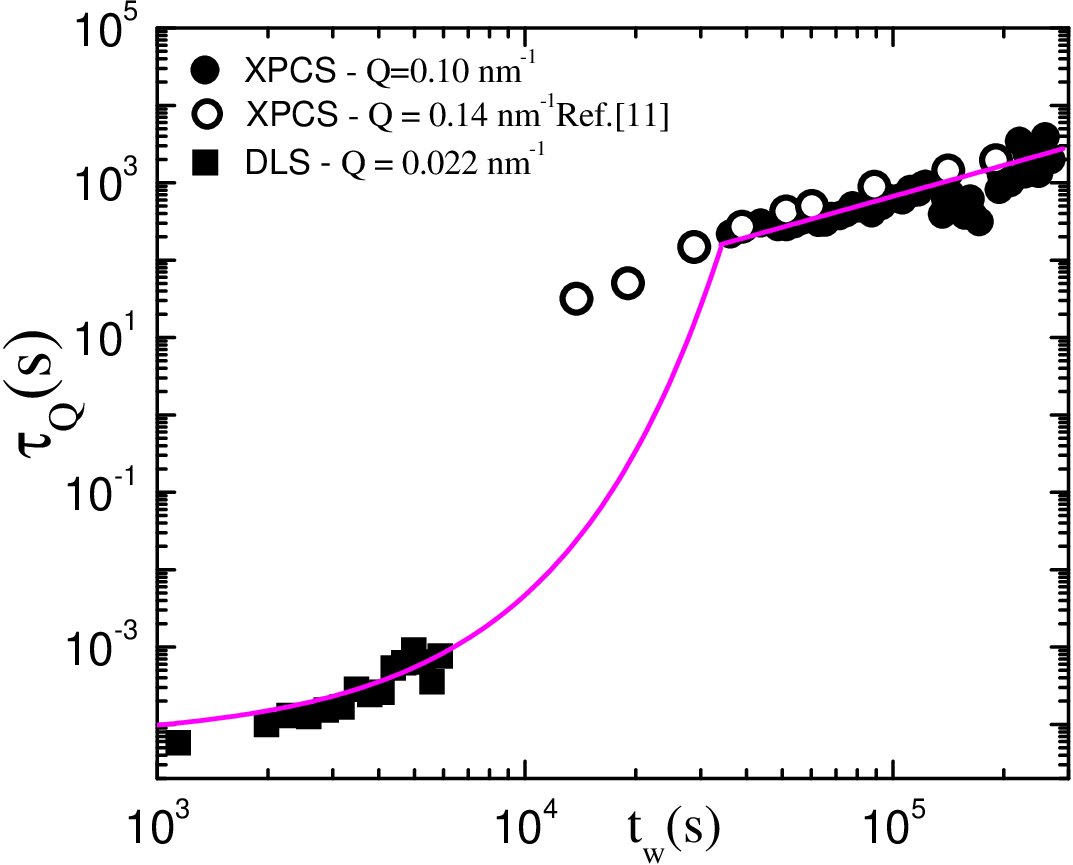}
\caption{Relaxation time as a function of the waiting time for an
aqueous Laponite suspension at concentration $C_w=3.0 \%$ as
measured by DLS at Q=0.022 nm$^{-1}$ (full squares) and XPCS at
Q=0.10 nm$^{-1}$(full circles). The data are compared with XPCS
measurements by Bandyopadhyay {\em et
al.}~\cite{BandyopadhyayPRL2004} at the same concentration (empty
circles). The full lines represent the best fits with an
exponential and a power law behaviour at small and large $t_w$,
respectively.} \label{Fig1}
\end{figure}

In summary, all the papers discussed above agree on the existence
of two regimes: at small $t_w$ an exponential aging regime with
$\tau_Q\sim exp(t_w/(s))$ and at larger $t_w$ a full aging regime
with $\tau_Q\sim t_w^{\alpha}$ and $1<\alpha<2$.

In this Communication we use XPCS and DLS to investigate the aging
dynamics of a colloidal system - Laponite suspensions - in the
glassy state~\cite{RuzickaPRL2010} at weight concentration
$C_w=3.0 \%$. The dynamical behaviour of spontaneously aged
samples has been compared to that of samples rejuvenated by the
application of a shear field after the glass formation. For
spontaneously aged samples a stretched behaviour with $\beta_Q<1$
is found in both aging regimes. Only rejuvenated samples show a
compressed behaviour with $\beta_Q>1$. The rejuvenation process,
i.e. the application of shear, allows to pass from a slower
(stretched) to a faster (compressed) than exponential behaviour.
In both cases for long waiting times the relaxation times scale as
$\tau_Q \sim Q^{-1}$. These results give further insights into the
topic of anomalous dynamics in soft materials, permitting also to
extrapolate analogies with structural glasses~\cite{RutaPRL2012}.

XPCS measurements were performed at the ID10A beamline at the
European Synchrotron Radiation Facility (ESRF) in Grenoble using a
partially coherent x-ray beam with a photon energy of 8 keV. A
series of scattering images were recorded by a charged coupled
device (CCD) and the ensemble averaged intensity autocorrelation
function $g_2(Q, t)=\frac{\langle\langle
I(Q,t_0)I(Q,t_0+t)\rangle_p \rangle}{\langle \langle
I(Q,t_0)\rangle_p \rangle}$, where $\langle...\rangle_p$ is the
ensemble average over the detector pixels mapping onto a single
$Q$ value and $\langle...\rangle$ is the temporal average over
$t_0$, was calculated by using a standard multiple $\tau$
algorithm~\cite{MadsenNJP2010,LehenyCOCIS2012}. To explore fast
time scales (t=$10^{-6}$ - 1 s), XPCS data were complemented by
DLS measurements performed at the ESRF light scattering laboratory
equipped with a Nd-YAG diode pumped laser (532 nm wavelength and
30 mW power), an avalanche photodiode and a logarithmic ALV
auto-correlator. The DLS measurements were performed at a
scattering angle $\theta=90^0$ which corresponds to a momentum
transfer $Q$ = 0.022 nm$^{-1}$. The combination of XPCS and DLS
permits to investigate exactly the same sample over a wide range
of $Q$, correlation times $t$ and waiting times $t_w$.

The samples, Laponite RD dispersions at weight concentration
C$_w=3.0 \%$, were prepared using the same protocol described
in~\cite{RuzickaSoftMat2011} which produces reliable and
reproducible samples. They were prepared and sealed in a glovebox
under $N_2$ flux to prevent CO$_2$ degradation. The powder,
manufactured by Laporte Ltd., was dispersed in pure deionized
water, stirred vigorously for 30 min, and filtered soon after
through 0.45 $\mu$m pore size Millipore filters. When dispersed in
water Laponite originates a charged colloidal suspension of disks
of  25 nm diameter and 0.9 nm thickness. Laponite dispersions are
usually considered monodisperse suspensions of single platelets.
However a grade of polydispersity has been found by different
authors~\cite{KroonPRE1996,BalnoisLangmuir2003}. The origin of the
waiting time ($t_w=0$) is the time at which the suspension is
filtered. Samples were placed and sealed in glass capillaries with
a diameter of 2 mm used for both XPCS and DLS measurements.

In Fig.~\ref{Fig1} the relaxation time $\tau_Q$ is reported as a
function of waiting time for a sample of C$_w$ = 3.0 $\%$. We
distinguish two different aging regimes in two distinct ranges of
aging times and for two Q vectors:  an exponential growth of the
relaxation times at early times $t_w$ measured by DLS at Q=0.022
nm$^{-1}$ and a power law behaviour at larger $t_w$ investigated
by XPCS at Q=0.10 nm$^{-1}$. These results are compatible with the
works discussed in the Introduction where a transition between
exponential and power law behaviours has been observed. At low
$t_w$ the DLS data (squares) have been obtained through the
standard fitting procedure which identifies a fast (almost age
independent) and a slow (strongly age dependent) relaxation
time~\cite{RuzickaPRL2004}. For these early ages, the slow
relaxation time $\tau_Q$ grows as $\tau_Q\sim exp(t_w/(s))$,  in
agreement with other light scattering
studies~\cite{CipellettiPRL2000, BellourPRE2003,SchosselerPRE2006,
RuzickaPRL2004}. For larger $t_w$ the fast relaxation time remains
out of the XPCS detection window, while the slow relaxation time
enters in the XPCS window at $t_w \sim 3 \cdot 10^4 s$. XPCS
measurements (full circles) at $Q$ = 0.10 nm$^{-1}$ show a
behaviour like $\tau_Q\sim t_w^{\alpha}$ with $\alpha = 0.94 \pm
0.07$. They are in good agreement with XPCS measurements performed
by Bandyopadhyay {\em et al.}~\cite{BandyopadhyayPRL2004} at Q =
0.14 nm$^{-1}$ and at the same concentration (empty circles).

\begin{figure}[t!]
\centering
\includegraphics[width=8cm,angle=0,clip]{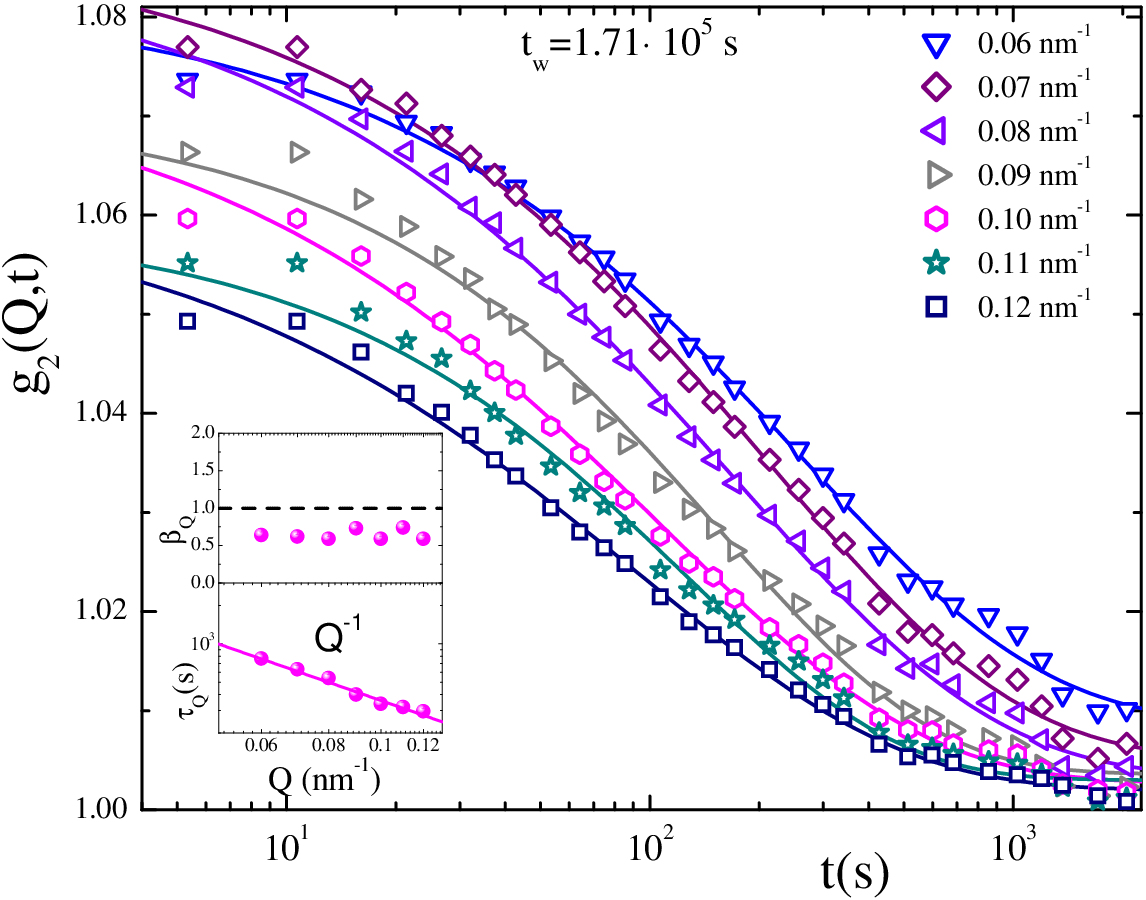}
\caption{Intensity autocorrelation functions of an aqueous
Laponite suspension at concentration C$_w$=3.0 $\% $, at waiting
time $t_w$=1.71 $\cdot 10^5$ s and at different Q values obtained
by XPCS (symbols). The solid lines represent the best fits
performed using Eq.~\ref{eq1}. Insets: $\beta_Q$ and $\tau_Q$
obtained from Eq.~\ref{eq1} as a function of Q. The full line in
the inset of $\tau_Q$ represents a fit with a power law.}
\label{Fig2}
\end{figure}

\begin{figure}[b!]
\centering
\includegraphics[width=8cm,angle=0,clip]{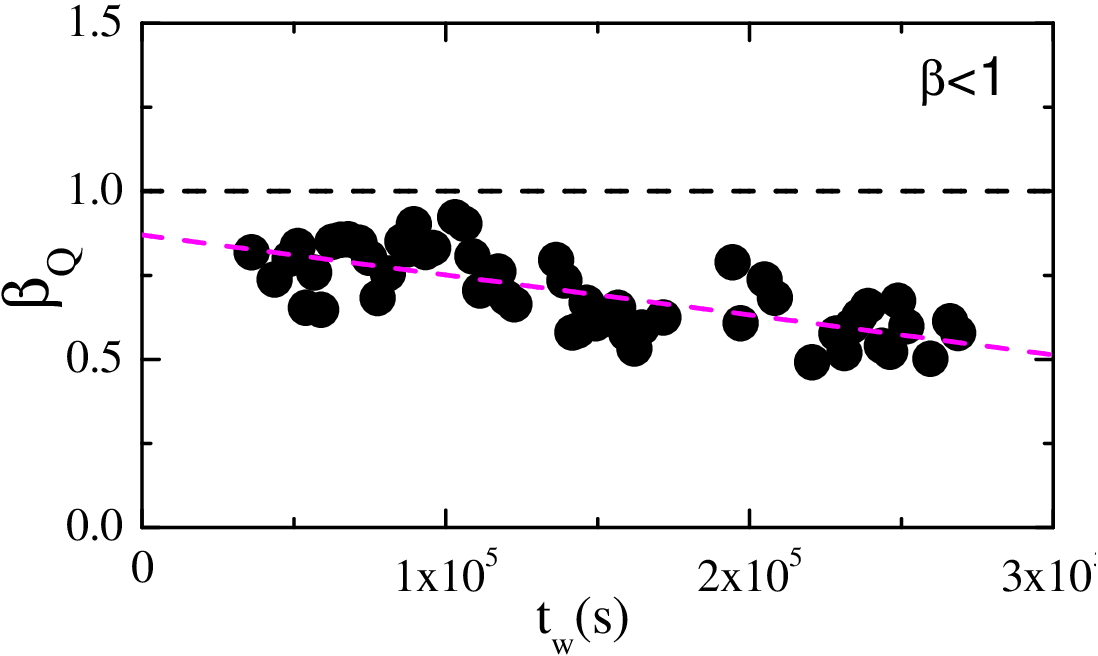}
\caption{$\beta_Q$ from Eq.~\ref{eq1} as a function of waiting
time for an aqueous Laponite suspension at concentration C$_w$=3.0
$\%$ and Q=0.10 nm$^{-1}$.} \label{Fig3}
\end{figure}

Figure~\ref{Fig2} shows, as an example, the intensity
autocorrelation functions of a Laponite sample at weight
concentration C$_w$=3.0 $\%$, after a waiting time $t_w$=1.71
$\cdot 10^5$ s and at different $Q$ values. For all the aging
times and $Q$ the XPCS data are well described by the fitting
expression:

\begin{equation}\label{eq1}
g_2(Q,t)= b[(Ae^{-(t/\tau_Q)^{\beta_Q}})^2+1]
\end{equation}

where $ b \cdot A^2 $ represents the contrast, $\tau_Q$ the
relaxation time and $\beta_Q$ the Kohlrausch exponent. The latter
two parameters characterize the microscopic dynamics of the
sample. The fits are shown as full lines in Fig.~\ref{Fig2}. The
relaxation time $\tau_Q$ vs $Q$, shown in the inset of
Fig.~\ref{Fig2}, displays a $\tau_Q \sim Q^{-1}$ behaviour which
is signature of a non free diffusive dynamics. The exponent
$\beta_Q$ is plotted vs $Q$ in the inset of Fig.~\ref{Fig2}. Its
values are always well below 1 for all the investigated $Q$ and at
all the investigated waiting times, as shown in Fig.~\ref{Fig3}
for Q=0.10 nm$^{-1}$ , indicating a stretched exponential
behaviour. This marks an important difference with respect to
previous studies of Laponite where compressed relaxations were
always observed in the full aging regime.

In summary, our measurements on spontaneously aged samples confirm
the well established exponential to power law crossover at
increasing waiting time but in addition two new important
observations are made: {\em i)} In contrast with previous studies,
a stretched behaviour ($\beta_Q<1$) is also found in the long time
regime, and {\em ii)} In the full aging regime a ballistic-type
behaviour ($\tau_Q \sim Q^{-1}$) is observed together with
stretched intensity correlation functions. The 1/Q dependence of
$\tau$ associated with a stretched exponential for the correlation
functions, is a surprising behavior that calls for theoretical
explanations and further experimental studies on different
systems. Furthermore these findings are in agreement with
theoretical~\cite{BhattacharyyaJCP2010} and
numerical~\cite{SciortinoPRE1996} results on supercooled liquids
where the decreasing temperature plays the same role as an
increasing waiting time in our system~\cite{RuzickaNatMat2011}.

\begin{figure}[t!]
\centering
\includegraphics[width=8cm,angle=0,clip]{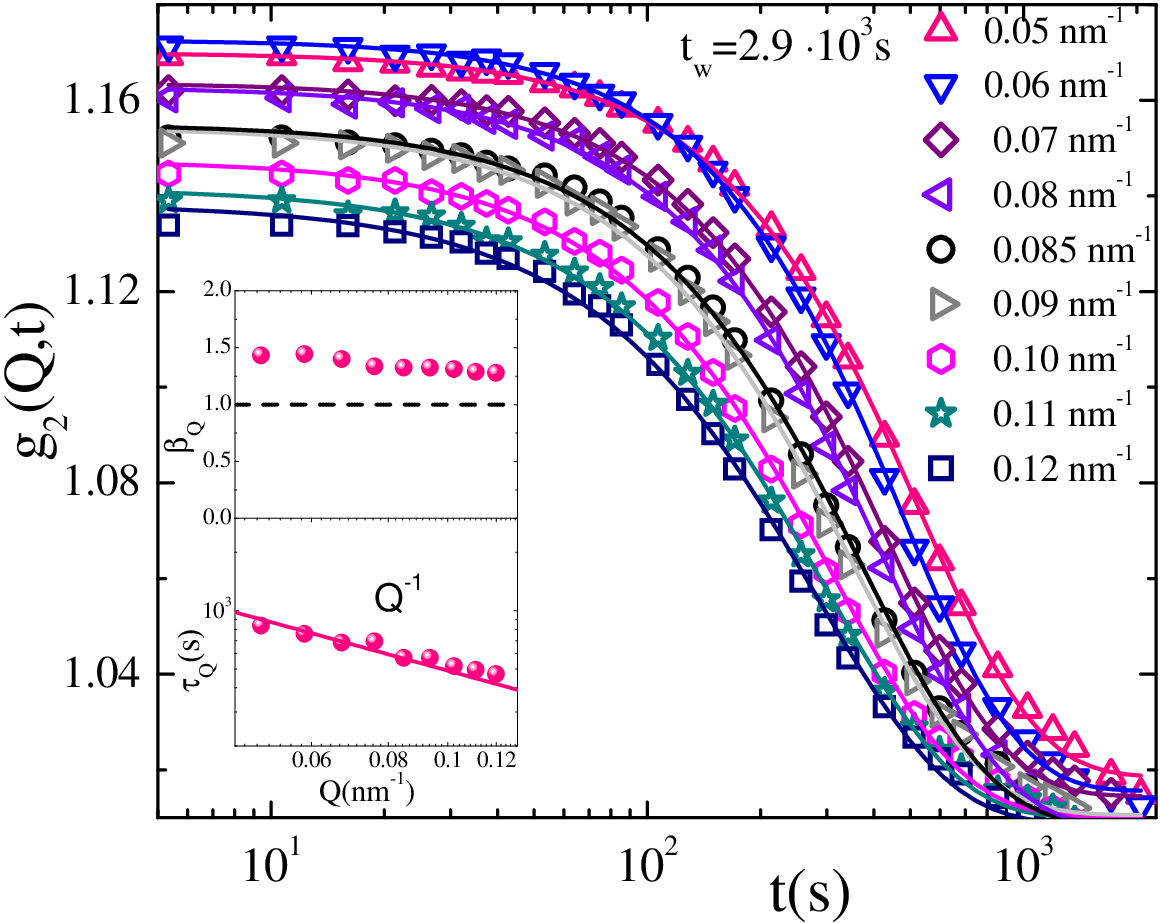}
\caption{Intensity autocorrelation functions of a rejuvenated
aqueous Laponite suspension ($t_R \sim $ 3.5 days) at
concentration C$_w$=3.0 $\% $, at waiting time
$t_w$=2.9$\cdot10^3$ s and at different Q values obtained through
XPCS (symbols). The solid lines represent the best fits performed
using Eq.~\ref{eq1}. Insets: $\beta_Q$ and $\tau_Q$ from
Eq.~\ref{eq1} as a function of Q. The full line in the inset of
$\tau_Q$ represents a fit with a power law.} \label{Fig4}
\end{figure}

\begin{figure}[b!]
\centering
\includegraphics[width=8cm,angle=0,clip]{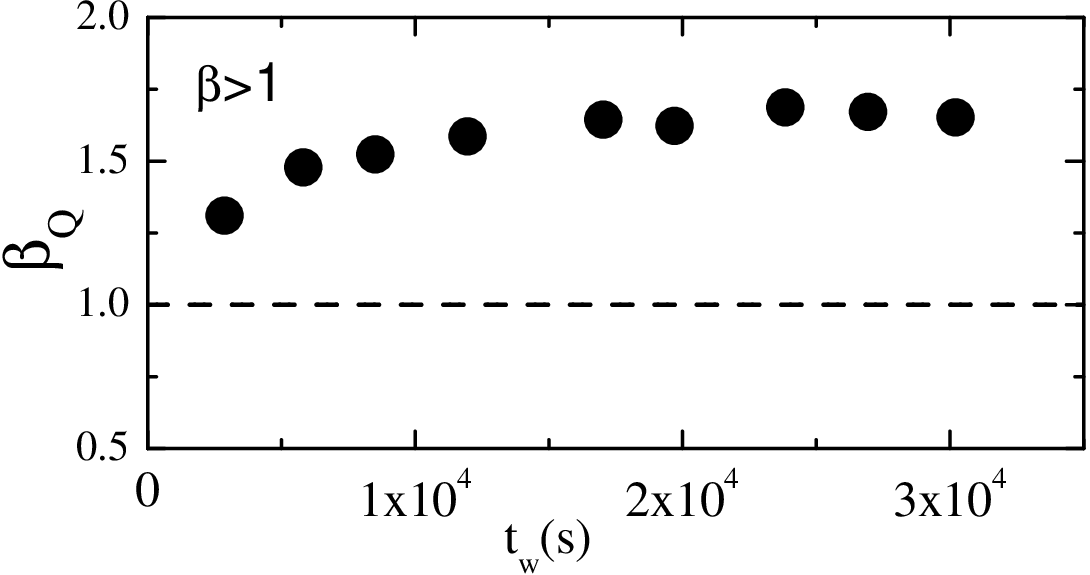}
\caption{$\beta_Q$ from Eq.~\ref{eq1} as a function of the waiting
time for an aqueous Laponite suspension at concentration
C$_w$=3.0$\%$ and Q=0.10 nm$^{-1}$  rejuvenated at $t_R \sim$ 3.5
days.} \label{Fig5}
\end{figure}

In order to reconcile our findings with the previous observations
on Laponite, we note that in
Ref.\cite{BellourPRE2003,SchosselerPRE2006} the measurements in
the full aging regime were performed after the sample, taken from
a stock solution, was injected in the sample container by a
syringe, thus applying a shear field to an aged sample. Therefore
we decided to study the aging dynamics of a sample rejuvenated by
a shear field which corresponds to returning it to earlier aging
times. The rejuvenation process never rewinds the sample to the
original as prepared one, in particular it has been observed that
the dynamical properties of Laponite strongly depend on the time
elapsed before rejuvenation~\cite{ShahinLangmuir2010}. Concerning
the structural properties we have verified that the static
structure factors of both spontaneously aged and rejuvenated
samples are those typical of a glassy Laponite
sample~\cite{RuzickaPRL2010}. The rejuvenation time $t_R$ is the
time elapsed between sample preparation and application of a shear
with a syringe on the arrested sample. For these rejuvenated
samples the origin of the waiting time $t_w=0$ is the moment of
application of the shear. In Fig.~\ref{Fig4} intensity
autocorrelation functions of a C$_w$=3.0$\%$ sample rejuvenated at
$t_R \sim 3.5$ days are shown. The data, in the $Q$-range between
0.05 nm$^{-1}$ and 0.13 nm$^{-1}$, are well described by the
fitting expression of Eq.~\ref{eq1} and the fits are plotted as
full lines in Fig.~\ref{Fig4}. The varying contrast with respect
to Fig.~\ref{Fig2} suggests that the fast dynamics  which is not
captured by the XPCS measurements~\cite{GuoPRE2010,
CzakkelEPL2011}, probably is different for rejuvenated and
spontaneously aged samples. The relaxation time $\tau_Q$ vs $Q$,
shown in the inset of Fig.~\ref{Fig4}, still carries the $\tau_Q
\sim Q^{-1}$ signature of non free diffusive
dynamics~\cite{MadsenNJP2010, LehenyCOCIS2012}. The exponent
$\beta_Q$ vs Q is reported in the inset of Fig.~\ref{Fig4} and
surprisingly its values are now always above 1 for all the
investigated $Q$ and for all the investigated waiting times, as
shown in Fig.~\ref{Fig5} for Q=0.10 nm$^{-1}$. This is markedly
different from the dynamics of the spontaneously aged sample that
is always stretched ($\beta_Q<1$ - Fig.~\ref{Fig3}).

The compressed exponential behaviour of the intensity
autocorrelation functions on the rejuvenated sample is in
agreement with the aforementioned experiments on Laponite
suspensions~\cite{BellourPRE2003,BandyopadhyayPRL2004,SchosselerPRE2006}.
This similar behaviour indicates that probably the samples of
Refs.~\cite{BellourPRE2003,BandyopadhyayPRL2004,SchosselerPRE2006}
have been prepared in a way that make them equivalent to
rejuvenated ones~\cite{EXPLAN}.

In the case of rejuvenated samples the shear applied by the
syringe induces internal stresses responsible for the compressed
exponential relaxations in agreement with several models
explaining this type of phenomenology in aging soft materials:
both a heuristic interpretation~\cite{CipellettiPRL2000} and a
microscopic model~\cite{BouchaudEPJE2001} underline the importance
of internal stresses in these complex systems. In particular
Ref.~\cite{CipellettiFD2003} proposes that the unusual slow
dynamics are due to the relaxation of internal stresses, built
into the sample at the transition to the arrested state. An
analogous interpretation can be hypothesized for most disordered
systems where internal stresses are induced by a quench from the
fluid to the glassy state. Recently, this has been experimentally
observed for a metallic structural glass~\cite{RutaPRL2012} where
the compressed behaviour was attributed to internal stress
relaxations. In this context, our experiment has shed new light on
this topic, showing that the same system can feature either
stretched or compressed exponential relaxations, depending on
whether it is spontaneously aged or rejuvenated. In this case the
internal stresses are induced in the rejuvenated sample by shear
(applied by the syringe) which plays the same role as temperature
quenching does in structural glasses.

In conclusion, through  XPCS and DLS  measurements, we have found
that spontaneously aged colloidal suspensions of Laponite at high
concentration $C_w=3.0\%$ are always characterized by a stretched
behaviour ($\beta_Q <1$) of the time correlation functions both in
the cage-forming and in the full aging regimes. A compressed
behaviour ($\beta_Q>1$) is found for the same sample but only
after rejuvenation. In both cases the relaxation times scale as
$Q^{-1}$ in the full aging regime. These results, addressing the
general topic of anomalous dynamics in soft materials, evidence a
novel dichotomic stretched/compressed dynamics in a colloidal
glass. The rejuvenation process, {\em i.e.} the application of
shear, allows to pass from a slower to a faster than exponential
decay. The clear correlation between rejuvenation by shear and the
occurrence of compressed exponential behaviour will help in
classifying past and future experimental observations and connect
with simulation studies and theoretical work.

We acknowledge ESRF for beamtime and Abdellatif Moussa{\"i}d for
providing assistance during the measurements. AF acknowledges
support from DOE grant EAC02-98CH10886 (NSLS-II project).


\providecommand*{\mcitethebibliography}{\thebibliography} \csname
@ifundefined\endcsname{endmcitethebibliography}
{\let\endmcitethebibliography\endthebibliography}{}

\end{document}